\documentclass{article}
\usepackage{amsmath}
\usepackage[numbers, square, sort&compress]{natbib}

\usepackage{hyperref}
\usepackage[top=2cm, bottom=3cm, left=2.5cm, right=2.5cm]{geometry}

\title{Note: A demonstration that classical gravity does not produce entanglement}
\author{Mike D. Schneider\footnote{Corresponding author; \href{mailto:schneidermd@missouri.edu}{schneidermd@missouri.edu}} \footnote{Department of Philosophy, University of Missouri} , Nick Huggett\footnote{Department of Philosophy, University of Illinois Chicago} , Niels Linnemann\footnote{Department of Philosophy, University of Geneva}}
\date{March 2026}

\begin{document}

\maketitle

\begin{abstract}
Once again, dispute has arisen over the interpretation of proposed quantum information theory experiments to probe the quantum nature of gravity by testing for gravitationally induced entanglement (GIE) between two spatially separated massive particles (\cite{aziz2025classical} vs. \cite{marletto2025classical, marlettoETC2025classical}; further contributions in \cite{diosi2025no,lin2025newtonian}). The confusion appears to reside in interpreting applications of a Hamiltonian formalism. But classical gravity cannot mediate entanglement on independent grounds. A Newton-Cartan analysis shows that if gravity is classical, a mediator, and entanglement is observed as an outcome of performing a GIE experiment, something other than gravity must have supplied the (virtual) force needed during the experiment to produce the effect. 

\end{abstract}

\section{Introduction}

In \cite{huggett2023quantum}, we historically situated and critically reviewed recent proposals, based on quantum information theory, to produce and observe ``gravitationally induced entanglement'' (GIE) to test whether gravity, construed as a mediator between spatially separated massive quantum systems in the laboratory, is itself quantum \citep{bose2017spin, marletto2017witnessing}. These proposals have generated a considerable literature (for extended discussion, see \cite{huggett2023quantum}), as well as active attempts to carry out such an experiment (but see \cite{Aspelmeyer} on the technical difficulties).

Confirmation of the quantum nature of gravity is argued by proponents to follow the successful execution of such an experiment if the (otherwise isolated) massive quantum systems arranged in spatial superpositions---`gravcats'---become entangled. The logic of this confirmation depends on a background claim from quantum information theory: 

\begin{itemize}
\item[\textsc{Claim}] Interactions with classical systems, particularly mediators governed by classical gravity, do not produce entanglement between quantum systems.
\end{itemize}

Recently, \cite{aziz2025classical} has disputed \textsc{Claim}, and thus disputed as well the conclusion that observing entanglement between the gravcats would confirm the quantum nature of gravity. To do so, they purport to show that classical gravity can produce entanglement by providing a particular Hamiltonian, which they interpret as generating quantum communication and therefore violating an underlying LOCC assumption familiar in quantum information theory (because it violates the CC part: classical communication). Likewise, in the original version (v.1) of \cite{lin2025newtonian}, the authors purported to show that classical gravity can produce entanglement by providing a particular Hamiltonian, which they interpret as generating non-local effects and therefore violating the same LOCC assumption (now because it violates the LO part: local operations). In fact, the latter invokes what we explicate in \cite{huggett2023quantum} as a naive ``Newtonian model'' of GIE: an idealization in which the experiment is described as a \textit{bipartite} system, with a direct, non-local interaction term between the gravcats, not a \textit{tripartite} system that includes a gravitational \textit{mediator}. So such a model is simply silent on any remark, like \textsc{Claim}, which assumes a gravitational mediator.\footnote{After posting the initial version of this note to the ArXiv, we learned that shortly prior, a subsequent version of \cite{lin2025newtonian} had reversed the authors' original pronouncement.  In the later version, the authors provide demonstrations in alignment with the standard reasoning that holds Newtonian gravity to be a mediator: ``We show that the mesoscopic quantum bodies with their parities of mass quadrupoles interacting via Newtonian gravity can get entangled only if the parity of the gravitational tidal field sourced by the quantum body is also quantized'' (v.2 abstract).}

\cite{aziz2025classical} (and \cite[v.1]{lin2025newtonian}) proceed by interpreting interaction Hamiltonians, and they accept a standard idealization that Newtonian gravity suffices to capture the role of gravity in the experiment protocol, rather than relativistic gravity. Objections to the original authors' analysis of the Hamiltonian they provide have already appeared \cite{marletto2025classical,marlettoETC2025classical,diosi2025no,di2025simple}. But classical gravity cannot produce entanglement on independent grounds, having nothing to do with a Hamiltonian formalism. In this note, we revisit existing analysis in \cite{huggett2023quantum}, noting that an immediate corollary demonstrates that in the agreed regime of the experiment, if classical \textit{gravity} is taken to be the Newtonian limit of classical Einstein gravity, then gravitational mediation simply will not induce entanglement. \textsc{Claim} is on solid grounds.

In what follows, we draw heavily on material that was previously relegated to an appendix in \cite{huggett2023quantum}. Our impression is that appendices can go unnoticed.

\section{A Newton-Cartan Analysis}

A central issue appears to be that working in a Hamiltonian formalism encourages taking the Newtonian limit to be a theory of a gravitational potential in flat, Galilean spacetime. This view seriously distorts the nature of gravitational mediation in General Relativity (GR), because it obscures the fact that gravity in GR is ultimately encoded in spacetime curvature, not a force field, and in this sense spacetime itself is the mediator. (In many applications, the choice of idealization of course does not matter, but if one is specifically concerned with the question of what can be achieved by classical, relativistic gravitational mediation in the Newtonian limit, then it matters a great deal.) 

Fortunately there is an alternative way of taking the limit, which yields Newton-Cartan gravity (\cite{cartan1923,cartan1924}). Newton-Cartan gravity (NCG) is arguably more faithful to the nature of general relativity (\cite{malament1986newtonian, christian1997exactly} and references therein), since it preserves the general relativistic insight that gravity manifests as dynamical geometry. Of course, the more usual flat spacetime formulation of gravity as a potential in the Newtonian limit can be obtained from the Newton-Cartan formulation, but NCG provides a more fruitful and physically appropriate setting to ask and answer foundational questions. This holds whether general relativity is considered on its own terms as a theory of gravity that is classical and relativistic (as in this note), or else it is considered as a pre-quantum theory of gravity, the states of which are assumed to offer approximate descriptions of coherent states of the quantum theory (as in \cite[Appendix B]{huggett2023quantum}, discussed shortly). 

As further evidence for (and explication of) the claim that NCG captures the Newtonian limit in a way relevant to the investigation into gravitational mediation, note that NCG can be quantized into a gauge field theory, whose quanta (``Newton-Cartan gravitons'') are (infinite speed) longitudinal bosons \cite{christian1997exactly}. These, like the more familiar gravitons of quantized linear GR, form a \textit{third} subsystem \textit{mediating} between a pair of massive subsystems. As we pointed out in  \cite{huggett2023quantum}, this tripartite subsystem picture is essential for sensibly invoking LOCC, so must be respected in any consideration of arguments that involve \textsc{Claim} articulated above. That is, the operative sense of `local' is that there is no direct interaction between the masses, but only with the mediator; indeed, our original point in invoking NCG in \cite[\S5.2.2]{huggett2023quantum} was to show explicitly that instantaneous propagation in a non-relativistic setting is not sufficient for mediating entanglement. That such a tripartite model arises from quantizing NCG---while there is \textit{no} corresponding quantization of standard Newtonian gravity as such a third mediating subsystem---is evidence that one can also safely take \textit{classical} NCG to capture the Newtonian limit in the most perspicuous way for investigating \emph{classical} mediation.\footnote{\cite{dibiagio2025circuitlocalityrelativisticlocality, di2025simple} argue that LOCC (and other no-go theorems) require a stricter sense of `mediator', which excludes gauge theories, including general relativity, and potentially NCG. We do not claim that our definition of `mediator' is sufficient for LOCC (though it is necessary), only that it is sufficient for a more general notion of mediation, in line with what is expressed by the impressionistic equation 2 in \cite{di2025simple}. Indeed, we take the discussion that closes \cite[\S F]{dibiagio2025circuitlocalityrelativisticlocality} as tacit acknowledgment that the more general notion is (sometimes) what is physically relevant.}

In NCG, as we have said, gravity is associated with spacetime curvature (as in GR). But the theory has the local symmetry structure of Galilean spacetime, not Lorentzian spacetime. In particular, time is absolute, moments correspond to spacelike hypersurfaces, and curvature vanishes at each moment (that is, non-trivial gravitational potentials only induce curvature in the direction of time). This theory is non-relativistic and  empirically indistinguishable from standard classical Newtonian gravity set in flat, Galilean spacetime that is given in terms of a potential (and corresponding force law). The models of NCG are represented as quadruples $(M,t_a,h^{ab},\nabla)$ where $M$ is a differentiable manifold, $t_a$ and $h^{ab}$ are tensorial quantities on $M$ such that $t_ah^{ab}=\textbf{0}$ for all $p\in M$, and $\nabla$ is a derivative operator on $M$ compatible with $t_a$ and $h^{ab}$. A standard procedure exists for relating $\nabla$ to a gravitational potential $\phi$ understood to supplement $(M,t_a,h^{ab},\mathring{\nabla})$, where $\mathring{\nabla}$ is flat. In all cases, derivative operators provide standards for sorting inertial (geodesic) trajectories. This is the sense in which the models of NCG replace models of different potentials by encoding gravity in spacetime curvature, analogous to the situation in GR but non-relativistic. 

In \cite[Appendix B]{huggett2023quantum} we reproduce the standard entanglement prediction of the GIE experiment protocol working with these non-relativistic spacetimes as the descriptions of the different branches of the wavefunction for the gravitational subsystem. That is, quantum states for the $\mathrm{gravcat}_1\times\mathrm{gravity}\times\mathrm{gravcat}_2$ system have the tripartite form $|\Psi_1\rangle\otimes|(M,t_a,h^{ab},\nabla)\rangle\otimes|\Psi_2\rangle$ and their superpositions, appropriate for a Newtonian limit of quantum gravity. Our original purpose in that appendix was to provide an analysis parallel to the derivation provided by \cite{christodoulou2019possibility}, which uses relativistic spacetimes as descriptions of the different branches of the wavefunction. The existence of the parallel analysis clarifies that the encoding of gravity as dynamical spacetime curvature alone suffices to generate the effect, once one moves into a quantum setting. That is, one does not need to invoke fundamentality arguments, appealing to relativistic locality, to produce entanglement. In other words: the assumption that gravity is a mediator---in line with the meaning of `local' in `LOCC'---and therefore that its mediating is what produces the entanglement effect, makes sense even starting from a manifestly non-local theory like Newtonian gravity, which in other formalisms (namely, a Hamiltonian formalism) includes a direct interaction term.

Whereas in the relativistic setting, the entanglement effect is recognized as a gravitational redshift, in the analysis of \cite[Appendix B]{huggett2023quantum} the entanglement effect is due to a `virtual force' term acting over the duration of the experiment, which makes up for a mismatch between different derivative operators relevant in describing different branches of the wavefunction within the experiment. The intuitive picture is that classically the gravitational field in the vicinity of one gravcat (considered as a test particle following worldline $\gamma\subset M$) depends on the separation from the other, which sources the field. Then, since different terms in the full wavefunction correspond to different separations, or different gravitational wavefunctions $|(M,t_a,h^{ab},\nabla)\rangle$ and $|(M,t_a,h^{ab},\nabla')\rangle$, each branch of the state corresponds to a different field for the gravcat. 

It turns out \cite[Appendix B, fn. 66]{huggett2023quantum}, that the difference between $\nabla$ and $\nabla'$ is equivalent to the influence of an ordinary force field that acts on $\gamma$ only in one or the other branch. That is, the effect on a particle of surveying an `errant' geometry $(M,t_a,h^{ab},\nabla')$ is captured by a force field in $(M,t_a,h^{ab},\nabla)$, acting on the particle.\footnote{\label{fntrivial}If $\nabla'=\nabla$ up to a constant multiple, corresponding to a gravitational potential $\phi'=\phi$ up to an additive constant, this force field trivializes to the $\textbf{0}$ tensor and zero work is put in. In this respect, $\nabla'$ is not errant, after all.} So, on the assumption that $\gamma$ is geodesic with respect to $\nabla$, in one of the models but not the other, work is required for the particle to follow that trajectory: one of the branches of the wavefunction involves (virtual) work for the particle to stick along $\gamma$. The upshot of this is that the work put in within one branch of the wavefunction and not the other is no more or less mysterious than that which is put into any charged particle in the presence of a fixed, ambient field only present in one branch, which deflects that charged particle off of its geodesic path in accordance with a force law associated with the field-charge pair.

Such a force differential between branches produces a phase shift between them, and entanglement between the gravcats, in the usual way. To calculate this shift one moves from work to total energy over the lifetime of the experiment, by integrating the scalar product of power and proper time over the curve $\gamma$. Thus, if $\xi^a$ is the four-velocity of the point mass $m$, here is the expression for the total energy input to the particle over the lifetime of the experiment, with respect to $(M,t_a,h^{ab},\nabla)$, which is required due to its surveying an errant $\nabla'$:

\begin{equation}
    \text{Total Energy} = m\int_\gamma{\xi^b(\xi^m\xi^n t_m t_n)\nabla_b(\phi'-\phi) (t_c\xi^c) dS}.
\end{equation}

$t_c\xi^c$ is the proper time, which in this setting necessarily agrees with the absolute time elapsed between the two endpoints of $\gamma$, the total lifetime of the experiment $T$. Given the static nature of the experiment protocol, the expression further simplifies as $m(\phi'-\phi)*T$,  where $(\phi'-\phi)$ is understood as a function solely of spatial coordinates at a point, corresponding to the difference between the gravitational potentials associated with $\nabla'$ and $\nabla$. (For full details, see the original work.) Finally, as is explicitly argued in that appendix, this quantity can be regarded as equivalent to the value of a classical action, which upon quantization yields a phase factor in agreement with existing analyses of the experiment. 

What is significant for present purposes, however, is that in the case where the underlying theory is not a quantum theory whose basis states are approximated as distinct NCG models centered on $\gamma$, and instead there is just one single model (whatever it is), everything in the preceding analysis goes through, but with $\nabla'=\nabla$. It is immediate that the phase factor trivializes (disregarding that computing the phase factor is arguably no longer physically motivated if the wavefunction branches are identical), and we can clearly say why: the relevant force field that showed up earlier in the analysis is the $\textbf{0}$ tensor (cf. footnote \ref{fntrivial}). To produce entanglement requires otherwise. We stress that although this observation about classical NCG mediation was absent in \cite{huggett2023quantum} due to the emphasis there on quantum NCG mediation, it follows immediately as a consequence.

Note that while we have argued that NCG is the appropriate Newtonian limit for the analysis of classical gravitational mediation (hence our demonstration that it induces no entanglement), there are outstanding questions regarding the applicability of our quantum Newton-Cartan derivation of the standard phase difference. First, one would like to recover this picture from the quantized NCG developed in \cite{christian1997exactly}, in analogy to the relativistic case \cite{chen2023quantum}. Second, there is the much harder question of whether quantized NCG as formulated in \cite{christian1997exactly} can be obtained as the Newtonian limit of some more plausible theory of quantum gravity than quantized linear GR, given that classical NCG is the limit of \textit{full} GR.

\section{Commentary}

An immediate consequence of results obtained in \cite[Appendix B]{huggett2023quantum} establishes that classical gravity, formulated as NCG, a manifestly non-local theory, cannot as a mediator induce entanglement. We stress that all parties should agree that, within the regime of the experiment protocol, this particular theory is the appropriate treatment of classical Einstein gravity. The use of the Newton-Cartan formalism just makes clear what is at stake in an antecedent commitment to think of classical gravity as a mediator (despite its non-local character as that mediator, given the experimental regime). 

What this means is that if entanglement is observed as an outcome of performing a GIE experiment, and if gravity is classical, something else is responsible for the entanglement. 

\section*{Acknowledgments} 

We thank Andrea di Biagio for helpful explications on the content of his note \citep{di2025simple}.

\bibliographystyle{abbrvnat}
\bibliography{finalbib}
\end{document}